# β-Ga$_2$O$_3$ NEMS Oscillator for Real-Time Middle Ultraviolet (MUV) Light Detection

Xu-Qian Zheng, *Student Member, IEEE*, Jaesung Lee, *Student Member, IEEE*, Subrina Rafique, Md Rezaul Karim, Lu Han, Hongping Zhao, Christian A. Zorman, *Senior Member, IEEE*, and Philip X.-L. Feng, *Senior Member, IEEE*

*Abstract*—We report on the first beta gallium oxide (β-Ga$_2$O$_3$) crystal feedback oscillator built by employing a vibrating β-Ga$_2$O$_3$ nanoresonator as the frequency reference for real-time middle ultraviolet (MUV) light detection. We fabricate suspended β-Ga$_2$O$_3$ nanodevices through synthesis of β-Ga$_2$O$_3$ nanoflakes using low-pressure chemical vapor deposition (LPCVD), and dry transfer of nanoflakes on microtrenches. Open-loop tests reveal a resonance of the β-Ga$_2$O$_3$ device at ~30 MHz. A closed-loop oscillator is then realized by using a combined optical-electrical feedback circuitry, to perform real-time resonant sensing of MUV irradiation. The oscillator exposed to cyclic MUV irradiation exhibits resonant frequency downshifts, with a measured responsivity of $\Re \approx$ -3.1 Hz/pW and a minimum detectable power of $\delta P_{min} \approx$ 0.53 nW for MUV detection.

## I. Introduction

PHOTODETECTION in the solar-blind ultraviolet (UV) regime ($\lambda$ < 280 nm) has received significant attention thanks to the absence of $\lambda$ < 280 nm light in the earth's atmosphere, providing a 'null' background in the solar-blind UV range, enabling ultrasensitive and selective illumination detection in missile tracking, fire detection, and environmental monitoring applications [1,2]. Beta gallium oxide (β-Ga$_2$O$_3$), a semiconductor with a monoclinic crystal structure, has an ultrawide bandgap, $E_g \approx$ 4.5–4.9 eV [3,4,5], perfectly aligned with the cutoff wavelength of the solar-blind range. Endowed with the ideal bandgap and promising electronic properties, photodetectors (PDs) made of crystalline β-Ga$_2$O$_3$ could capitalize on optoelectronic interactions in the solar-blind UV range [6,7,8]. In addition, owing to its excellent mechanical properties (*e.g.*, Young's modulus, $E_Y \approx$ 260 GPa) [9], β-Ga$_2$O$_3$ is an attractive and promising structural material for innovating nano/microelectromechanical systems (NEMS/MEMS) and new transducers, especially sensors that can benefit from its ultrawide bandgap. Given their ultrahigh responsivities to external stimuli, and enhanced by scaling, resonant-mode NEMS are engineered as novel physical sensors, *e.g.*, infrared detectors [10]. To date, thin films of β-Ga$_2$O$_3$ have been mainly fashioned into electronic devices without any moving element, such as field effect transistors (FETs) [11,12,13] and diode UV PDs [6,7,8]. Transducers that exploit the mechanical properties of β-Ga$_2$O$_3$ are yet to be demonstrated and carefully studied.

In this Letter, we describe the construction and measurement of the first self-sustaining β-Ga$_2$O$_3$ crystal oscillator by using a β-Ga$_2$O$_3$ resonator and a feedback circuitry, for middle ultraviolet (MUV, 200–300 nm, strongly overlapping with the solar-blind UV range) light detection and real-time sensing. We first characterize the open-loop responses of the β-Ga$_2$O$_3$ resonator; we then devise and implement an optical-electrical feedback circuitry, to realize the oscillator operations, and demonstrate real-time sensing of cyclic MUV irradiation on the device. Fig. 1 illustrates the light sensing mechanism of the β-Ga$_2$O$_3$ nanomechanical resonator. The photothermal effect induced by the incident MUV light elevates the temperature and expands the suspended β-Ga$_2$O$_3$ crystal, leading to a resonance frequency downshift. By probing the resonance frequency shift of the β-Ga$_2$O$_3$ device, the intensity of the incident MUV light can be resolved to achieve MUV detection.

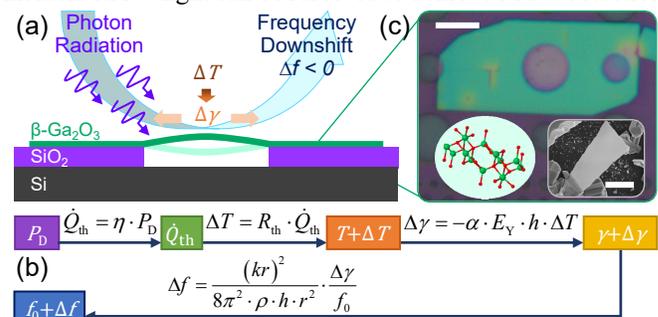

Fig. 1. (a) Illustration of the MUV light sensing mechanism. (b) Signal transduction chain analysis. $P_D$: irradiation power on device; $R_{th}$: thermal resistance; $\dot{Q}_{th}$: rate of heat flow; $\eta$: optical absorption coefficient; $T$: temperature; $\gamma$: surface tension; $\alpha$: thermal expansion coefficient; $E_Y$: Young's modulus; $h$: thickness; $r$: radius; $\rho$: mass density; $(kr)^2$: eigenvalue; $f_0$: resonance frequency. (c) β-Ga$_2$O$_3$ nanomechanical device image. *Insets*: The crystal structure of β-Ga$_2$O$_3$ and a scanning electron microscopy (SEM) image of a β-Ga$_2$O$_3$ nanoflake on the growth substrate. Scale bars: 5 μm.

## II. Device Fabrication and Characterization

Device fabrication consists of the synthesis and transfer of β-Ga$_2$O$_3$ nanoflakes for making suspended nanostructures. We

X.-Q. Zheng, J. Lee, C.A. Zorman, and P.X.-L. Feng are with Department of Electrical Engineering & Computer Science, Case Western Reserve University (CWRU), Cleveland, OH 44106 USA (e-mail: philip.feng@case.edu).
S. Rafique, L. Han and H. Zhao were with Department of Electrical Engineering & Computer Science, CWRU, Cleveland, OH 44106 USA.
M. R. Karim, and H. Zhao are with the Department of Electrical and Computer Engineering, Ohio State University, Columbus, OH 43210 USA.

perform low-pressure chemical vapor deposition (LPCVD) using a 3C-SiC-on-Si substrate as a template for β-Ga$_2$O$_3$ nanoflake synthesis. Using high purity Ga pellets and O$_2$ gas precursors in a 950 °C environment for 1.5 hours, the formation of β-Ga$_2$O$_3$ nanostructures proceeds, step by step, from β-Ga$_2$O$_3$ nanocrystals to nanorods, and then to nanoflakes via extrusion, without the need of any foreign catalyst [14].

We fabricate suspended β-Ga$_2$O$_3$ nanomechanical structures from β-Ga$_2$O$_3$ nanoflakes by using a dry transfer technique [9]. Fig. 1(c) shows a suspended β-Ga$_2$O$_3$ device made of a 73nm-thick β-Ga$_2$O$_3$ nanoflake. The flake is suspended over a ~5.2 μm-diameter circular microtrench. We use this device to build the β-Ga$_2$O$_3$ oscillator for MUV detection.

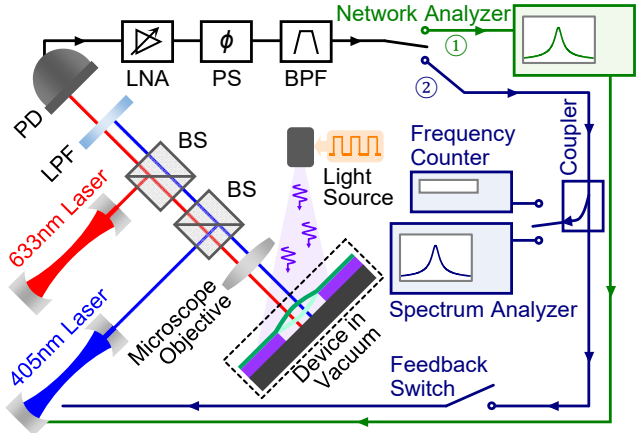

Fig. 2. Schematic of real-time resonance tracking system for MUV light detection utilizing the feedback oscillator built by using a β-Ga$_2$O$_3$ resonator as the frequency reference. PD: photodetector; LPF: long-pass filter; BS: beam splitter; LNA: low-noise amplifier; PS: phase shifter; BPF: band-pass filter.

## III. RESULTS AND DISCUSSIONS

Fig. 2 illustrates the real-time resonance tracking system utilizing the self-sustaining β-Ga$_2$O$_3$ feedback oscillator for MUV light sensing. We use a home-built ultrasensitive laser interferometry system to measure both the undriven thermomechanical resonance and the photothermally driven response of the β-Ga$_2$O$_3$ resonator in moderate vacuum (~20 mTorr). We use a 633nm laser to interferometrically resolve the device motion, which is detected by a sensitive PD. For detecting the undriven thermomechanical motion, only the 633nm laser is employed, with the signal PD output being fed to a spectrum analyzer [9,15,16]. In the driven response measurement (path ① in Fig. 2), a network analyzer is used to modulate the intensity of an amplitude-modulated 405nm laser in order to photothermally drive the nanoflake, and then detect the device motion using the 633nm laser through the PD [16].

Further, a closed-loop oscillator can be constructed by enabling the feedback circuit in path ② of the experimental system. The amplitude and phase of the signal from the PD are adjusted by tuning the amplifier and phase shifter, respectively, to satisfy the Barkhausen criterion. After proper filtering, the signal is sent to modulate the 405nm laser, thus converted from electrical back to optical domain again. In this configuration, the mechanical motion is self-sustaining without an external ac driving force, establishing a feedback oscillator. A spectrum analyzer and a frequency counter are used to record the oscillator output spectrum and monitor the frequency (Fig. 2).

We employ a continuous light source to illuminate light onto the device down to the MUV regime (200–300 nm), for real-time characterization of the MUV light sensing performance of the β-Ga$_2$O$_3$ resonator embedded in the loop (Fig. 2). By switching the light source on and off and monitoring the oscillator output, the oscillator frequency response to the MUV irradiation can be resolved and recorded.

We first measure the resonance frequency of the β-Ga$_2$O$_3$ nanomechanical resonator. Using path ①, we measure an open-loop, photothermally driven resonance of the device at ~30 MHz with a quality factor ($Q$) of ~150 (Fig. 3(a)). Using path ② with the feedback switch off, the resonator also exhibits a measured thermomechanical resonance at ~30 MHz with a $Q \sim 200$ (Fig. 3(b)), as read out by the spectrum analyzer.

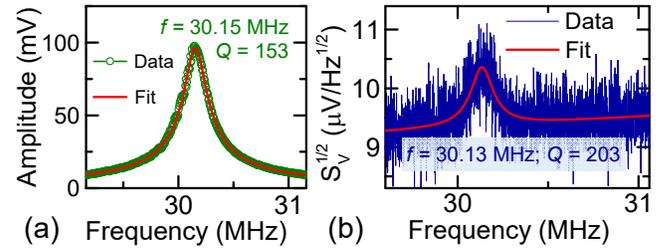

Fig. 3. (a) Photothermally driven response of a circular drumhead β-Ga$_2$O$_3$ resonator. (b) Thermomechanical noise spectrum of the resonator.

To enable real-time MUV sensing, we build a self-sustaining oscillator by connecting the resonator to a combined optical-electrical feedback circuitry (Fig. 2, path ② with the feedback switch on). We first characterize the oscillator in the frequency domain by using a spectrum analyzer. With the oscillator, the $Q$ of the ~30MHz resonance is boosted from ~200 to an effective $Q_{\text{eff}} >15,000$ (Fig. 4(a)), which is a >70-fold enhancement. We also measure the oscillator frequency fluctuations and compute the Allan deviation, to evaluate the frequency stability of the oscillator (Fig. 4(b)), which yields Allan deviation of $\sigma_A (\tau) \sim 3.9 \times 10^{-5}$ for $\tau = 10$ ms.

With the $E_g \approx 4.5$–$4.9$ eV bandgap of crystalline β-Ga$_2$O$_3$, photons with wavelength below 280 nm could be efficiently absorbed by β-Ga$_2$O$_3$. We investigate the MUV light detection characteristics of the β-Ga$_2$O$_3$ oscillator by illuminating light containing MUV photons onto the device. The power intensity from the aforementioned light source is adjusted to be $p_S = 0.5$ and 1 W/cm$^2$. Therefore, the incident power of the light on device ($P_D$) can be calculated using $P_D = A_D \cdot T_{\text{op}} \cdot (d_S^2/d_C^2) \cdot p_S$, where $A_D$ is the device area, $T_{\text{op}}$ is the optical transmission coefficient of the chamber optical window, $d_S$ and $d_C$ are the diameters of the circular MUV light source and the light spot on device substrate, respectively. Thus, we have incident power levels of ~24 nW and ~49 nW on the suspended circular drumhead. The results of real-time tracking of the oscillator frequency show clear frequency downshifts upon cyclic illumination down to MUV range (Fig. 5). Using $\Re = \Delta f/P_D$, we obtain a responsivity of $\Re = -3.1$ Hz/pW. Given the Allan deviation, the oscillator has a frequency fluctuation of $\delta f (\tau = 10$ ms) = $(2)^{1/2} \sigma_A f_0 \approx 1.66$ kHz. Thus, the minimum detectable power (MDP) of the MUV sensing oscillator is $\delta P_{min} = \delta f/\Re \approx$ 1.4 nW at $\tau$ = 1 s, and $\delta P_{min} \approx$ 0.53 nW at $\tau$ = 10 ms, *i.e.*, better resolution in a higher speed detection. This clearly demonstrates an intrinsic advantage of feedback oscillators for real-time, high speed sensing.

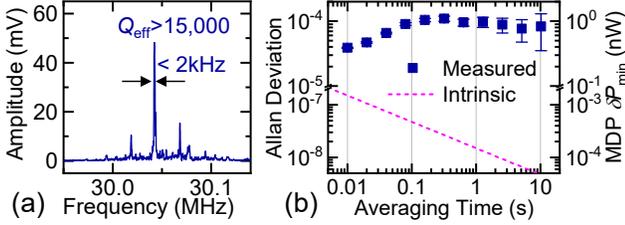

Fig. 4. Performance of the β-Ga$_2$O$_3$ feedback oscillator. (a) The closed-loop oscillation spectrum. (b) Measured and intrinsic Allan deviation.

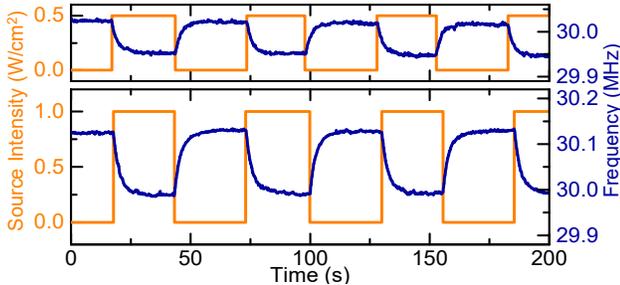

Fig. 5. Responses of the β-Ga$_2$O$_3$ oscillator to photon irradiation. Note the base frequency has a slight drift between two separate measurements (1W/cm$^2$ cyclic data taken ~30 mins earlier), likely due to adsorbates or other drifting effects.

Further, to analyze the fundamental limit of the β-Ga$_2$O$_3$ resonator and oscillator for MUV detection, we calculate its thermomechanical noise limited Allan deviation, $\sigma_{A,th}(\tau) = (\pi k_B T/(P_c \tau Q^2))^{1/2}$, where $k_B$ is the Boltzmann constant, $T$ is temperature, and $P_c$ is the operating power of the resonator [17]. We have $P_c = 4\pi^3 f_0^3 M_{eff}(0.745 a_c)^2/Q$, where $a_c$ is the critical amplitude $a_c = (1.54/(\beta Q))^{1/2}$ [18], $\beta = (23-9\nu)(1+\nu)/(56h^2)$ [19], $\nu = 0.2$ is the Poisson's ratio, and $M_{eff} = 0.1828 \rho h \pi r^2$ is the effective mass for the first mode. Thus we get $\sigma_{A,th}(\tau = 10\text{ ms}) = 1.50 \times 10^{-7}$, corresponding to a frequency noise of $S_f^{1/2}(f) = 2(\pi\tau)^{1/2} f_0 \sigma_A = 1.60$ Hz/Hz$^{1/2}$ and a noise equivalent power NEP = $S_f^{1/2}/(2\pi|\Re|) = 8.2 \times 10^{-14}$ W/Hz$^{1/2}$. The much better intrinsic Allan deviation proves that the oscillator performance can be improved through future engineering. Table I shows metrics benchmarking of this work with previously reported resonant UV detectors, including thin film bulk acoustic resonators (FBAR) and surface acoustic wave (SAW) devices using ZnO or AlGaN [20,21,22]. The β-Ga$_2$O$_3$ oscillator shows much better performance in both $|\Re|$ and MDP. In addition (see Table II), the β-Ga$_2$O$_3$ oscillator also exhibits similar NEP and better MDP compared to conventional β-Ga$_2$O$_3$ UV PDs [6,7,8].

The measured frequency downshifts upon MUV irradiation could be explained by light absorption and photothermal effects. With the $E_g \approx$ 4.5–4.9 eV bandgap, β-Ga$_2$O$_3$ crystal will absorb MUV photons and elevates its lattice temperature. The suspended β-Ga$_2$O$_3$ crystal, which has a positive thermal expansion coefficient ($\alpha = 1.5-3.4 \times 10^{-6}$ K$^{-1}$ [23]), expands, thus 'softening' the resonator and causing a resonance frequency downshift. The substrate underneath the β-Ga$_2$O$_3$ resonator consists of a SiO$_2$ layer on top of bulk Si. The bandgap of SiO$_2$ (~9 eV) is too wide to contribute to photothermal absorption of the device structure. Rather, SiO$_2$ layer serves as a mechanical isolator against the thermal expansion of the MUV-absorbing Si bulk. Further, the low thermal conductivity of SiO$_2$ ($\kappa_{SiO2}$ = 1.4 W/(m·K)) also makes it a good heat barrier between the β-Ga$_2$O$_3$ device ($\kappa_{\beta-Ga2O3}$ = 10–27 W/(m·K) [24]) and the Si substrate ($\kappa_{Si}$ = 148 W/(m·K)). With the relatively low thermal conductivity of β-Ga$_2$O$_3$ and the considerable thermal resistance between β-Ga$_2$O$_3$ and SiO$_2$ due to van der Waals contact, the photothermal heating is efficiently confined in the β-Ga$_2$O$_3$ suspended structure, ideally facilitating the MUV sensing. Accordingly, the β-Ga$_2$O$_3$ nanomechanical feedback oscillator could provide excellent potential for future MUV, hence solar-blind UV ($\lambda <$ 280 nm), detection, and the performance can be enhanced by further engineering of device structure and integration with circuit.

TABLE I. COMPARISON WITH UV SENSING RESONANT TRANSDUCERS

| Active Area | Device Type | $|\Re|$ (Hz/nW) | MDP, $\delta P_{min}$ | Refs. |
|---|---|---|---|---|
| 21.2 μm$^2$ | β-Ga$_2$O$_3$ NEMS | 3,100 | 0.53 nW | This Work |
| 0.026 mm$^2$ | ZnO FBAR | 63 | 6.5 nW | [20] |
| 1.6 mm$^2$ | ZnO SAW | 2.3 | n/a | [21] |
| n/a | AlGaN SAW | 0.5 | n/a | [22] |

TABLE II. COMPARISON WITH β-Ga$_2$O$_3$ PHOTODETECTORS

| Active Area | Responsivity $|\Re|$ | MDP$^a$ $\delta P_{min}$ | NEP$^b$ | Refs. |
|---|---|---|---|---|
| 21.2 μm$^2$ | 3.1 Hz/pW | 0.53 nW | 8.2×10$^{-14}$ W/Hz$^{1/2}$ | This Work |
| 0.8 cm$^2$ | 39.3 A/W | 28.0 nW | 1.5×10$^{-14}$ W/Hz$^{1/2}$ | [6] |
| ~7 mm$^2$ | 0.07 A/W | 1–10 nW | ~8×10$^{-14}$ W/Hz$^{1/2}$ | [7] |
| ~0.8 mm$^2$ | 8.7 A/W | 1–10 nW | ~7×10$^{-15}$ W/Hz$^{1/2}$ | [8] |

$^a$MDPs for PDs are calculated here, by using $\delta P_{min} = I_D/\Re$; $I_D$ is dark current.
$^b$NEPs for PDs are calculated here, by using NEP = $(2qI_D)^{1/2}/\Re$, where $q$ is electronic charge, $\Re$ is responsivity, and $2qI_D$ is shot noise of the PD.

IV. CONCLUSION

This Letter presents the first demonstration of a self-sustaining β-Ga$_2$O$_3$ crystal feedback oscillator by employing a vibrating β-Ga$_2$O$_3$ nanomechanical resonator (at ~30MHz) as its frequency reference. The feedback oscillator is further employed to demonstrate real-time sensing of cyclic MUV light irradiation onto the β-Ga$_2$O$_3$ resonator. This study reveals the potential of using β-Ga$_2$O$_3$ NEMS as ultrasensitive detectors for solar-blind UV ($\lambda <$ 280 nm), and using β-Ga$_2$O$_3$ feedback NEMS oscillators for real-time MUV sensing, which could lead to future applications, including target acquisition, flame detection, and environmental monitoring.

ACKNOWLEDGMENTS

We thank the US Department of Energy (DOE) EERE (Grant DE-EE0006719), the Army Research Office (ARO) (Grant W911NF-16-1-0340), the National Science Foundation (NSF) SNM Award (Grant CMMI-1246715), and the ThinkEnergy Fellowship (X.-Q. Zheng) for financial support. S. Rafique, M. Rezaul Karim and H. Zhao thank the NSF DMR Program (Grant DMR-1755479) for financial support.